\documentclass[twocolumn,pra,article,showpacs,showkeys,10pt]{revtex4-1} 
\usepackage{amsmath,amssymb,graphicx}
 \usepackage{natbib}
 \usepackage[titletoc]{appendix}
\usepackage[colorlinks, citecolor=blue,linkcolor=blue]{hyperref}
\usepackage{hyperref}
\usepackage{epstopdf}
\begin{document}
\title{Assessing the Effects of Macroeconomic Variables on Child Mortality in D-8 Countries Using Panel Data Analysis}
\author{M. Waseem Akram$^1$}
\email{waseemakram727@yahoo.com}
\affiliation{$^1$Department of Economics, The Islamia University of Bahawalpur, 64100, Bahawalpur, Pakistan}
\author{Binita Shahi$^2$}
\email{b.shahi@students.rave.ac.uk}
\affiliation{$^2$Department of Management Sciences, Ravensbourne University, SE10 0EW, London, United Kingdom}
\author{M. Javed Akram$^{3*}$}
\email{javed.akram.16@alumni.ucl.ac.uk }
\affiliation{$^3$Department of Physics and Astronomy, University College London, Gower Street, London WC1E 6BT, United Kingdom}

\begin{abstract}
This research analyses the axiomatic link among health expenditures, inflation rate, and gross national income (GNI) per capita concerning the child mortality (CMU5) rate in D-8 nations, employing panel data analysis from 1995 to 2014. Utilising conventional panel unit root tests and linear regression models, we establish that education expenditures, in conjunction with health expenditures, inflation rate, and GNI per capita, display stationarity at level. Additionally, we examine fixed effects and random effects estimators for the pertinent variables, utilising metrics such as the Hausman Test (HT) and comparisons with CCMR correlations. Our data demonstrate that the CMU5 rate in D-8 nations has steadily decreased, according to a somewhat negative linear regression model, therefore slightly undermining the fourth Millennium Development Goal (MDG4) of the World Health Organisation (WHO).
\end{abstract}
\maketitle 
\section{Introduction}
The fourth Millennium Development Goal (MDG4) of the World Health Organisation (WHO) was to decrease under-5 child mortality (U5CM) by two-thirds between 1990 and 2015 (\textcolor{blue}{WHO}). These endeavours have facilitated the reduction of under-five mortality (U5M) to an estimated 7.6 million in 2010, a decrease from the 9.6 million reported in 2000, and the first time it has fallen below 10 million (\textcolor{blue}{Liu, Johnson, \& Cousens 2010}). Although progress has been achieved, recent analyses indicate that only 10 of the 67 countries with high child mortality ($\ge$ 40 deaths/1000 live births) are on the right track to meeting the MDG-4 target. In contrast, developing countries such as Sub-Saharan Africa continue to face this challenge \textcolor{blue}{$^2$}.

In this context, recent decades have seen a substantial spectrum of research examining the correlation between child mortality rates (CMRs) and various characteristics, including parental education, observed globally \textcolor{blue}{(Adamou et al. 2024)}. A multitude of research have shown the correlation between U5CM and the educational attainment of mothers and parents. Additionally, IMR has been the subject of other factors. Typhoid and its severe medication resistance have significantly contributed to human mortality. It is projected that over 10 million deaths are attributable to complexity and drug-resistant threats to humanity, therefore presenting a significant problem for policymakers about health expenditures. The NHS in the UK expends £1 million annually \textcolor{blue}{(Ward et al., 2025)}. This investment far exceeds the one allocated to developing nations such as Pakistan and India annually. Conversely, economic development (ED) is no longer confined to a continuous rise in certain criteria, such as gross national income (GNI), while disregarding others. Additional elements are now seen as fundamental components of this process; the most significant being enhancements in health, particularly infant mortality rates (IMRs) and child mortality rates (CMRs), advancements in education, and management of inflation, among others. The health of children and adolescents is a paramount global health concern, with infant mortality rates (IMR) and child mortality rates (CMR) regarded as key indicators of a nation's health status (\textcolor{blue}{Wang 2002},\textcolor{blue}{Wang et al. 2025}).

Since the foundational research by Preston in 1975 and subsequently by Pritchett and Summers in 1996, it has been established that affluent individuals tend to be healthier, as evidenced by life expectancy and child mortality rates within nations \textcolor{blue}{(Preston 1975), (Pritchett and Summers, 1996)}. Furthermore, a strong correlation exists between higher national income and improved health outcomes for the population of that country. A recent report by \textcolor{blue}{(Liang et al. 2018, Adamou et al. 2024)} indicates that mortality rates in neonates (under 28 days) and infants (under 1 year) are disproportionately elevated, accounting for over two-thirds of deaths in children under 5 years of age \textcolor{blue}{(Rajaratnam et al. 2010)}. Ninety percent of these fatalities transpire in developing nations \textcolor{blue}{(Ehret et al. 2017)}, with the most elevated newborn mortality rates seen in the resource-limited countries of sub-Saharan Africa and South Asia \textcolor{blue}{(Childinfo.org: 2013, Shandra et al. 2011)}, and low-income countries \textcolor{blue}{(Karunarathne et al.  2025)}.

Substantial advancements have occurred in the last twenty years regarding the study of child mortality from the perspectives of environmental and development economics (\textcolor{blue}{You et al. 2015, Wang et al. 2025}). In addition to numerous studies from a medical science perspective aimed at enhancing maternal and child health, there are significant health care interventions, such as skilled birth attendants and measles vaccination, alongside measures addressing socioeconomic inequities in middle- to low-income countries (\textcolor{blue}{Karunarathne et al.  2025}). Furthermore, there is an emphasis on expanding the coverage of high-impact child survival interventions, including insecticide-treated bed nets, nutritional support, care-seeking behaviour, and the treatment of childhood diarrhoea and pneumonia. The first category, significantly shaped by community infrastructure, encompasses: drinking water source, toilet facility type, and sewage system (\textcolor{blue}{Mejia 2024}). Recent metaphysical critiques have argued that both time and history lack independent ontological status, with significant implications for free will and political philosophy (\textcolor{blue}{Zaman, 2025}). The second category is significantly impacted by the home's socio-economic status; it encompasses flooring material, population density, cleanliness of the surrounding environment, and the cohabitation of farm animals and household members. The extent of this effect across various locations and the role of other socioeconomic issues, such as income inequality, has been a topic of continuous discussion in the literature. Numerous scholars have examined the correlation between income and child mortality via econometric techniques; nonetheless, the precise nature of this link remains unresolved (\textcolor{blue}{Mejia 2024, Fotio et al. 2024}).

In previous years, maternal education has been identified as a crucial factor influencing child mortality rates in developing nations \textcolor{blue}{Ragina Fuches at all (2010)}. Emily et al. have demonstrated (\textcolor{blue}{Smith-Greenaway et al. 2013}) the influence of mothers' formal education at the primary level on child mortality rate. Their studies indicate that reading skills contribute to understanding the association in Nigeria. Data from the demographic and health survey was utilised.They believe that women's reading skills improve significantly in relation to primary education. Which demonstrates that low reading skills play a vital role in child mortality. The data utilised comes from the 2003 Nigeria Demographic and Health Survey (NDHD), encompassing a sample of 12,076 children born to 4,576 women during the period from 1993 to 2003 (\textcolor{blue}{Adebayo et al 2005, Smith-Greenaway et al. 2013}).

Various donors and development organisations, in conjunction with the United Nations and national governments worldwide, pledged to attain a two-thirds decrease in under-five death rates from 1990 to 2015, as specified in the UN Millennium Declaration \textcolor{blue}{(Muhamad et al., 2025)}. Two critical measures for evaluating progress towards this target are the under-five mortality rate (U5MR) and the infant mortality rate (IMR) (UN Development Group, 2003). The main aim of this study is to record and assess the impact of education in D-8 nations ((\textcolor{blue}{Latif, 2025}). D-8, or Developing-8, is an association dedicated to promoting developmental collaboration among Bangladesh, Egypt, Indonesia, Iran, Malaysia, Nigeria, Pakistan, and Turkey. The D-8 was formally proclaimed via the Istanbul Declaration at the Summit of Heads of State/Government on June 15, 1997. The objectives of the D-8 Organisation for Economic Cooperation encompass improving the global economic status of member nations, broadening and creating new trade possibilities, augmenting participation in international decision-making processes, and raising living standards. D-8 signifies a worldwide framework rather than a localised endeavour, as seen by the variety of its member membership. The Organisation for Economic Cooperation (D-8) functions as a platform that does not undermine the bilateral and multilateral obligations of its member states, stemming from their associations with other international or regional entities  (\textcolor{blue}{Iqbal et al. 2016, Iqbal et al. 2023, Shah et al. 2020, Khalid et al. 2023, Selcuk et al. 2025, Balsalobre-Lorente 2025}).
\section{Research Methods}\label{sec2}
\subsection{Data and variables}
Panel Data is a dataset that captures the behaviours of entities across time, where these entities may include states, wages, child mortality rates, education expenditures, health expenditures, GNI per capita, and inflation, among others. This type of data is referred to as {\it longitudinal} or {\it cross-sectional time-series data} (\textcolor{blue}{Maddala 2001, Lo Bue 2024)}. A panel data set, or longitudinal data set, typically comprises a time series for each cross-sectional unit within the dataset, as demonstrated in the current work. Before empirically estimating the panel model, it is crucial to evaluate the stationarity of the data from the D-8 nations; failing to do so increases the likelihood of obtaining erroneous findings (\textcolor{blue}{Campbell 1991}). Numerous tests exist to evaluate the stationarity of data, including the Augmented Dickey-Fuller (ADF) test (\textcolor{blue}{Davidson and MacKinnon, 2004}), the Levin, Lin \& Chu (LLC) test (\textcolor{blue}{Levin, Lin \& Chu 2002}), the Im, Pesaran and Shin W-statistics test \textcolor{blue}{(Im, K.S. et al. 2003)}, the Breitung T-statistics test \textcolor{blue}{(Breitung, 2001, Breitung and Das, 2005)}, and the Phillips and Perron test (\textcolor{blue}{Phillips, P.C.B., 1988}) . The aforementioned test will determine whether the data is stationary. In the current work, we employ the renowned Levin, Lin, and Chu (LLC) test \textcolor{blue}{(Levin, Lin, and Chu, 2002)} for unit root analysis at the stationary level. Additionally, we utilise fixed and random effects models, as well as the Hausman test (\textcolor{blue}{Hausman 1978}). Furthermore, we utilise the Pearson correlation coefficient test to examine the links between the dependent variables and the CM rate \textcolor{blue}{(Pearson, E.S., 1931)}.

Panel data analysis (PDA) is an extensively used statistical method leading to applications in numerous fields of research including for example  epidemiology, social science and econometrics in order to analyse two-dimensional -- typically cross sectional and longitudinal -- panel data (\textcolor{blue}{Maddala 2001}). The data are usually collected over time and over the same individuals and then a regression is run over these two dimensions. Multidimensional analysis is an econometric method in which data are collected over more than two dimensions (typically, time, individuals, and some third dimension) (\textcolor{blue}{Davies and Lahiri 1995}).

In this paper, we consider U5CM rate in D-8\textcolor{blue}{$^{3}$} countries as annual discrete time-series data (1994-2014), which is sourced from the database available at \textcolor{blue}{www.indexmundi.com}.
For the present panel study, we are interested to investigate the association among the dependent variable i.e. \textit{child mortality rate} (CMR) versus the dependent variables which are: (i) Education Expenditures, (ii) Health expenditures, (iii) Rate of inflation, and (iv) GNI per capita.
\subsection{The model: panel analysis}
A typical panel data regression model can be written as (\textcolor{blue}{Hsiao and Lahiri 1999}):
\begin{equation}
Y_{it} = \alpha + \beta X_{it} + \varepsilon _{it},
\end{equation}
where $Y$ is the dependent variable, $X$ is the independent variable, $\alpha$ and $\beta$ are coefficients, where $i$ and $t$ represent individuals and time. The last term is the error ${\displaystyle \varepsilon _{it}}$ which is often very important in the analysis. Generally, assumptions about the error term determine whether one should investigate either 'fixed effects' or 'random effects. In a fixed effects model, ${\displaystyle \varepsilon _{it}}$ is assumed to vary non-stochastically over 
$i$ or $t$ making the fixed effects model analogous to a dummy variable model in one dimension. In a random effects model,  ${\displaystyle\varepsilon _{it}}$ is assumed to vary stochastically over i or requiring special treatment of the error variance matrix.
\subsubsection{General form of the Model}
We model the child mortality (CM) in D-8 countries as a function of discrete time-dependent variables for the period of 1995-2014. For our case, Eq.~(1) takes the general form written as
\begin{equation}\label{eq1}
Y=f _t(E,H, G, I),
\end{equation}
where,  in equation (\ref{eq1}): (i) the variable $E$ represents the education expenditures in D-8 countries, whereas (ii) $H$ accounts for the Health expenditure, (iii) $G$ corresponds to gross national income (GNI) (per capita), and finally (iv) $I$ owns the inflation rate as fourth variable, whose dependence on CM is explored.
In order to represent the combined impact of four variables as defined above, we define the economic model for CM in D-8 countries as:
\begin{equation}\label{eq2}
C.M = \alpha_0 + \beta_1 (E) + \beta_2(H) + \beta_3(G) + \beta_4(I),
\end{equation}
where in the above model, $\alpha_0$ is a constant , the coefficients $\beta_i$ ($i=1-4$) represents the impact of: (1) $\beta_1(E)$ = Education expenditures, (2) $\beta_1(H)$ = Health expenditure, (3) $\beta_3(G)$ = gross national income (GNI) (per capita), and (4) $\beta_4(I)$ = inflation, respectively, on the CM in D-8 countries.

Panel data analysis has three more-or-less independent approaches: independently pooled panels; random effects models;
fixed effects models or first differenced models.
The selection between these methods depends upon the objective of the analysis, and the problems concerning the exogeneity of the explanatory variables.

\subsection*{(i) Stationary Analysis}
This section examines the stationarity of each dependent variable. Panel regressions need that variables be stable or cointegrated to prevent false outcomes. Consequently, we utilise the Levin, Lin \& Chu (LLC) Test (\textcolor{blue}{Levin, Lin \& Chu, 2002}) for the common unit root procedure. In \textcolor{blue}{Table-1}, we present the results of process of stationary of the concerning data in D-8 countries. We found that the stationary of data of the variables do not exist at level in table 1. 

This $T$-statistics tells about the data is while the column $P$-value is showing there probability values. The unit root tells us about the variables are stationary or non-stationary because of their probability value. In this study all variables are stationary.

The essence of the application of Levin-Lin-Chu Test (LLC) are: (i) \textbf{Null Hypothesis} - 
$H_0: \delta = 0$, which suggests that panel data is unit root (non-stationary data), whereas (ii) \textbf{Alternative Hypothesis}
$H_1: Low p-value \delta < 0.05)$: Reject null Hypothesis: variable is stationary, which reflects that panel data is not unit root (i.e. data is stationary).

Table-I allows us to validate the null hypothesis that the data exhibits a unit root by the conventional $t$ test. Nearly all variables exhibit stationarity at their levels. We see that the inflation rate is marginally acceptable in applied macro panel analysis, with a p-value ranging from $0.05$ to $0.09$. Therefore, proceeding with panel regression is warranted. Regrettably, we cannot proceed since the t-test is only relevant when the underlying time series is stationary. Alternatively, we may implement a test conducted by statisticians utilising fixed-effect and random-effect statistical methodologies, such as Stata and E-Views. SPSS and Origins software typically produce results in both tabular and graphical formats. The tau analysis is referred to as the Fixed-effect test (FET) in this domain. If data stationarity occurs at a certain level, it is determined by their probability value, indicating that the variables are stationary at that level. 

Furthermore, panel unit root testing originated from time series unit root testing. The primary distinction in time series testing for unit roots is the necessity to account for the asymptotic behaviour of the time-series dimension $T$ and the cross-sectional dimension $N$. The manner in which $N$ and $T$ approach infinity is essential for ascertaining the asymptotic behaviour of estimators and tests employed for non-stationary panels. Multiple methods have been documented for addressing the asymptotics \textcolor{blue}{(Robert Kunst's summary derived from Chapter 12 of Baltagi)} about panel unit root tests.

\begin{table}[ht]
\caption{Panel Unit Root Test Results}  
\centering          
\begin{tabular}{c c c}    
\hline\hline\\[0.02ex] 

Variables name & T-statistics & p-value \\[0.2ex]  

\hline    
Child mortality & -3.2815 & 0.0005 \\
Education Expenditures  & -2.5653 & 0.0052\\
Health Expenditures & -7.3085 & 0.0000\\
GNI per capita & -3.3089 & 0.0005\\
Inflation & -1.3760 &0.0844\\[0.5ex]  
\hline          
\end{tabular}
\label{table2}    
\end{table} 
\subsection*{(ii) Fixed Effects Estimator (within estimator)}
In \textcolor{blue}{Table-II}, we assess whether country-specific effects are random as opposed to fixed. The fixed effect estimator, akin to a first difference, serves as a modification to eliminate the unobserved random impact before estimating. There are several methods to remove unobserved heterogeneity; one such method is to consider a single exogenous model as:
\begin{equation}
y_{it} = \beta_{1}X_{it} + a_i + u_{it},
\end{equation}

\begin{table}[ht]
\caption{Panel Fix-effect test result}  
\centering          
\begin{tabular}{c c  c c}    
\hline\hline\\[0.02ex] 

Variables name & Coefficient  & T-test & Prob. value \\[0.2ex]  

\hline    
Child mortality & 20.94003  & -15.31  & 0.000 \\
Education Expenditures  & .0515033 & 0.03 & 0.978 \\
Health Expenditures & .0021814 & 3.74 &  0.000 \\
GNI per capita &  .1340339 & 2.24 & 0.026 \\
Inflation & 515.0225 & 18.28 & 0.000\\[0.5ex]  
\hline          
\end{tabular}
\label{table2}    
\end{table} 
where t is time period (1994-2014) ($t = 1, 2, ..., T$) and $i = 1, 2, ..., N$, $N$ is  number of observations.
We see that results are very similar to fixed effects. Notably, education and income reduce child mortality, whereas Inflation rate increases child mortality. Moreover, health spending remains insignificant.
\subsection*{(iii) Random Effect Model}
Alternatively, the random effects results are shown in \textcolor{blue}{Table III}, which indicate that probability values of all explanatory variables are statistically significant at $1\%$ level of significance and all coefficients shows positive relationship among dependent and all independent variables.

\begin{table}[ht]
\caption{Panel Random-effect test result}  
\centering          
\begin{tabular}{c c  c c}    
\hline\hline\\[0.02ex] 

Variables name & Coefficient  & T-test & Prob. value \\[0.2ex]  

\hline    
Child mortality & -21.21041 & -15.33 & 0.000\\
Education Expenditures  & .4736304 & 0.25 & 0.800 \\
Health Expenditures & .0021031 & 3.55 & 0.000 \\
 GNI per capita  & .1274093 & 2.11 & 0.035 \\
Inflation & 520.3752 & 17.21 & 0.000\\[0.5ex]  
\hline          
\end{tabular}
\label{table2}    
\end{table} 

\begin{equation}
C.M = \beta_{0} + \beta_{1} X_{it_1} + \beta_2 X_{it_2} + ... + \beta_{k} X_{it_k} + a_I + u_{it}
\end{equation}

Perhaps the most fundamental difference between fixed and random effect is of inference. A fixed-effects analysis can only support inference about the group of measurements (subjects, etc.) you actually have the actual subject pool you looked at. A random-effects analysis, by contrast, allows you to infer something about the population from which you drew the sample. If the effect size in each subject relative to the variance between your subjects is large enough, you can guess (given a large enough sample size) that your population exhibits that effect.

\subsection{Hausman Test}
To decide between fixed or random effects you can run a Hausman test (\textcolor{blue}{Hausman 1978, Nourin et al. 2025}) where the null hypothesis is that the preferred model is random effects vs the alternative the fixed effects (see Green, 2008, chapter 9). Hausman (\textcolor{blue}{Hausman 1978}) provided a specification test that differentiates between the fixed effects estimator and the random effects estimator. The Hausman test is designed to facilitate the decision between fixed and random effects estimations. The Hausman test is based on a straightforward concept which may also be employed to distinguish between fixed effects and random effects models in panel data  (\textcolor{blue}{Baltagi 2008b}). In this scenario, Random Effects (RE) is favoured under the null hypothesis due to its superior efficiency, however under the alternative, Fixed Effects (FE) is at least consistent and hence favoured. The Hausman statistic is developed based on the disparity between the two estimators. The sample distribution of the Hausman statistic delineates the threshold at which a difference becomes incompatible with the null hypothesis of accurate specification. The Hausman test is utilised to determine the optimum approach between fixed and random effects estimators. Primarily, it assesses whether the unique errors ($\delta_{i}$) are connected with the explanatory factors, with the null hypothesis positing that there is no correlation (\textcolor{blue}{Athey 2017}).
\subsection{Hypothesis of Hausman test}
\subsubsection{Null hypothesis}
The random effects model (REM) is suitable, when the following equation is suitable (\textcolor{blue}{Hausman 1978}):
\begin{equation}
H_0: Cov(U_{it}, X_{it}) = 0
\end{equation}
\subsubsection{Alternative Hypothesis}
Fixed effects model is suitable, when we have
\begin{equation}
H_1: Cov(U_{it}, X_{it}) \neq 0           
\end{equation}

\begin{table}[ht]
\caption{Results of Hausman test}  
\centering          
\begin{tabular}{c c  c c}    
\hline\hline\\[0.02ex] 

Variables name & Coefficient  & T-states & Prob. value \\[0.2ex]  

\hline    
Dependent variable: &~ & ~  &~\\
Child mortality  & 515.0225 & 18.28 & 0.000 \\
Independent variables: &~ & ~  &~\\
Education Expenditures &  - 20.94003 & -15.31 & 0.000\\
Health Expenditures & .0515033 & 0.03 & 0.978\\
GNI (Per capita) &  .0021814 & 3.72 & 0.000\\
Inflation  & .1340339 & 2.24 & 0.026\\[0.5ex]  
\hline          
\end{tabular}
\label{table2}    
\end{table} 

In case of Null hypothesis ($H_0$), consistency is present both in fixed and random effects. On the other hand, in case of Alternative hypothesis ($H_1$) only $\beta_{FE}$ is consistent. Consequently, if the difference between $\beta_{FE}$ and $\beta_{RE}$ is large then we reject $H_0$. Moreover, the acceptance or rejection of the criteria is based on the following conditions:  (i) We accept $H_0$ if: {\it Probability value} $> 5\%$. (ii) We reject the $H_0$ if: {\it Probability value} $< 5\%$. The above probability values are less than $5\%$, so we drop the $H_0$, which reflects that fix effect is appropriate to describe the results present in this paper (\textcolor{blue}{Asteriou and Hall, 2021}).
\section{Results and Discussions}
\subsection*{(i) Statistics of child mortality (CM) in D-8 countries and dependent variables}
\begin{figure}[ht]
\includegraphics[width=0.5\textwidth]{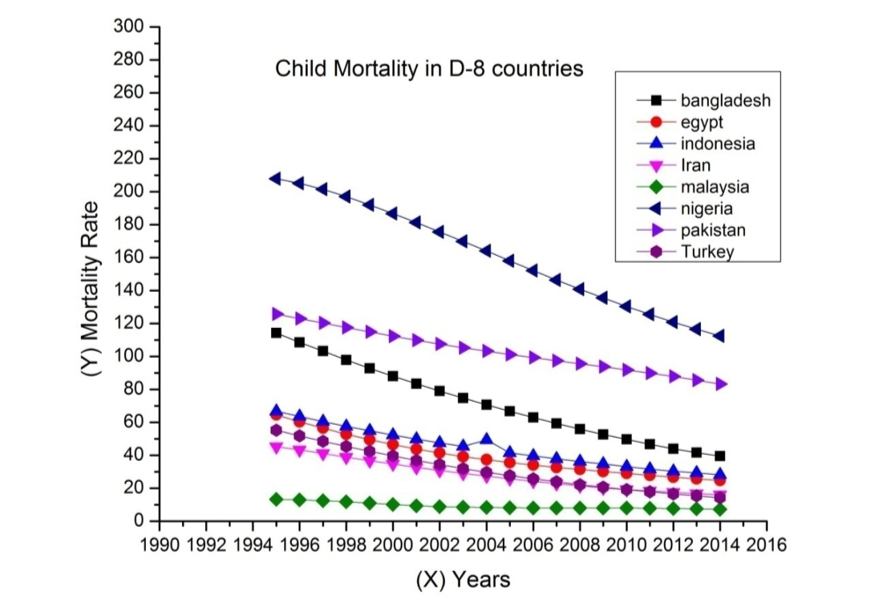}
\caption{Trends of mortality rate (CMU5) in D-8 nations: Data indicates that certain countries exhibit a pronounced reduction in mortality rates, which are consistently declining yearly at a rate more rapid than that of other nations.}\label{fig1}
\end{figure}
\subsubsection{Formula}
\begin{figure*}[ht]
\includegraphics[width=0.48\textwidth]{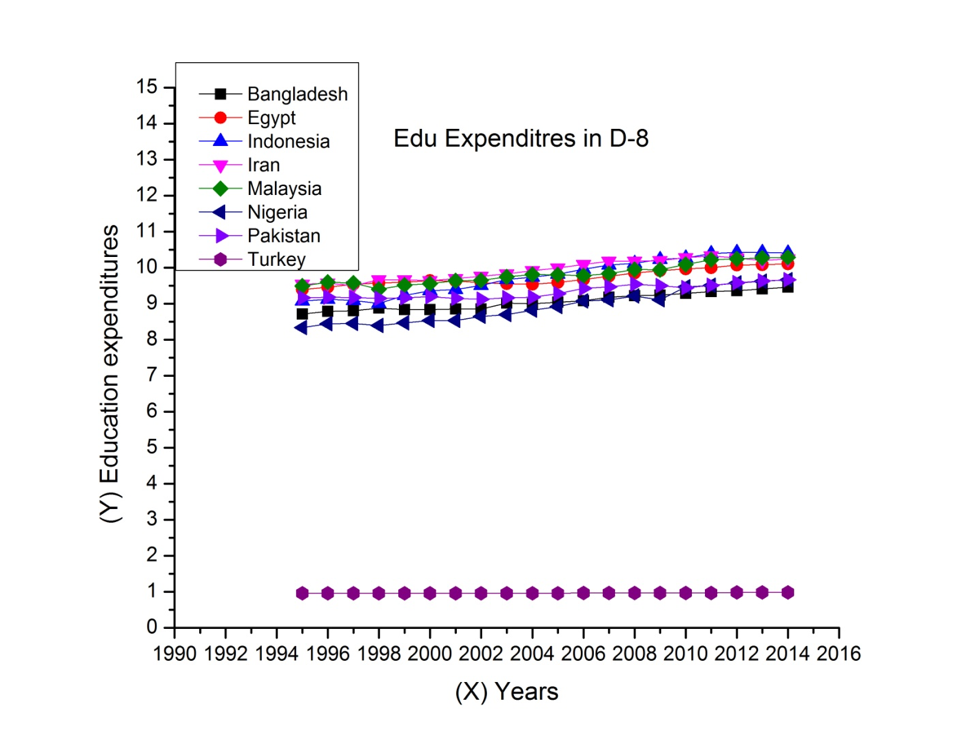}
\includegraphics[width=0.48\textwidth]{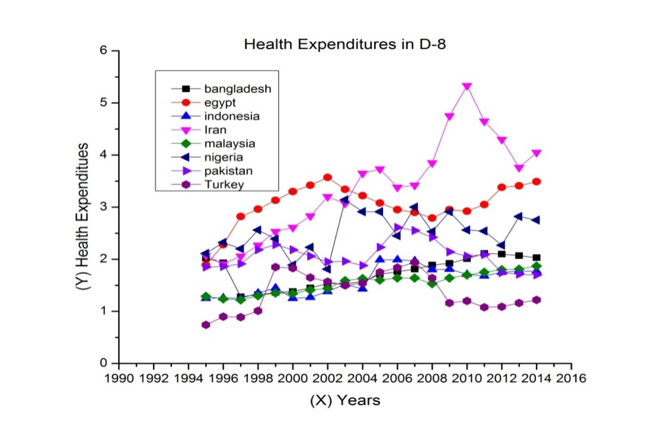}
\includegraphics[width=0.48\textwidth]{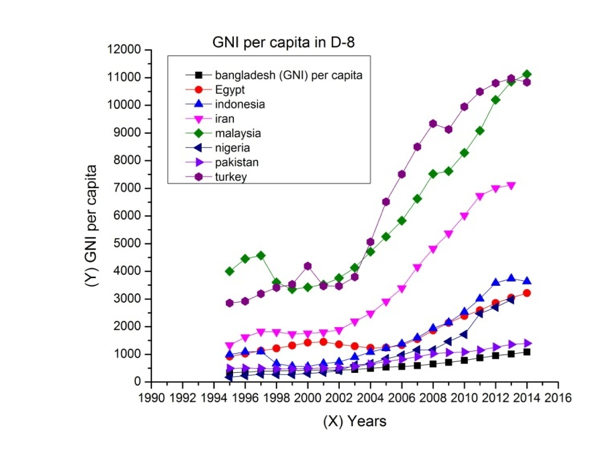}
\includegraphics[width=0.48\textwidth]{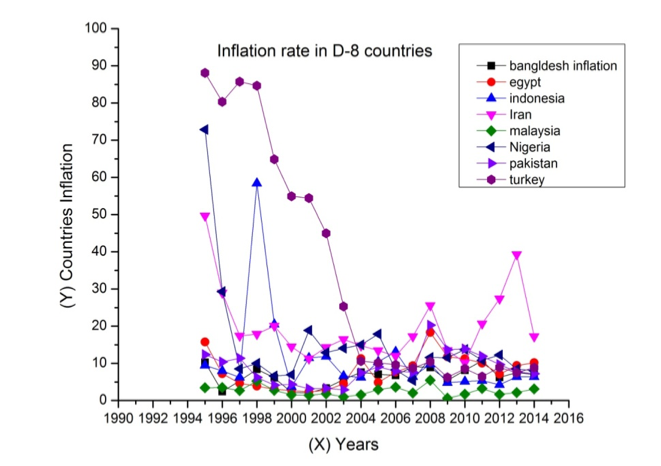}
\caption{Trends of independent variables in D-8 countries for our discrete time-series data: (a) Health expenditures, (b) Education expenditures, (c) GNI per capita, and (d) Inflation rate, respectively.}\label{fig2}
\end{figure*}

The child morality rate (CMR is given by,
\begin{equation}
CMR =\frac{D}{N} (1000),
\end{equation}
where, $D$ represents  number of deaths between 0-4  years during the period under consideration, and $N$ live birth of newborns during the year of calculation.
\subsection*{(ii) Education Expenditures (EE) in D-8 countries}
Government expenditure on education (current, capital, and transfers) is represented as a percentage of GDP. It encompasses expenditures financed by transfers from overseas sources to the government. General governance often denotes local, regional, and central authorities. To mitigate child mortality, increasing educational expenditures is optimal; yet, in Pakistan, only 2$\%$ of GDP is allocated to this sector. \textcolor{blue}{Figure 2A} illustrates the educational costs in D-8. Turkey must augment its educational spending, particularly in comparison to neighbouring countries. This aspect is crucial for decreasing child mortality.
Minor increments of these expenditures are not control child mortality as a pleasant manner.
\subsection*{(iii) Health Expenditures (HE) in D-8 countries}
Health spending in D-8 nations is shown in \textcolor{blue}{Figure 2B}. Total health spending includes direct and tax. Recurrent and capital spending make up direct expenditure. All health products and services excluding health capital are included in recurrent spending. It relies on the amount and distribution of these expenditures across the country. Any marginal increase in public sector health spending may boost human capital, economic growth, and child mortality, depending on the human resource situation \textcolor{blue}{(Maruthappu, Ng, Williams, Atun,\& Zeltner, 2015}).

\subsection*{(iv) GNI per capita in D-8 countries}
D-8 nations' GNI per capita shown in Figure 2C. Additionally, the development aims to boost GNP. Our approach improves D-8 nations' criteria to characterise development. This article investigates how D-8 GNI (per capita) skyrockets, reducing child mortality slowly. According to the data, education, health, GNI (per capita), and inflation all show modest improvement. In D-8 nations, education, health, inflation, and GNI (per capita) are strongly connected with child mortality. However, GNI (per capita) in D-8 is rising and infant mortality is falling, while all other indicators are behind. Bangladesh is one of the few nations on track to meet MDG 4 on child mortality. In recent years and among children over one month, mortality has dropped rapidly. This lowering rate may not be sustainable and may stop MDG 4 from being achieved. Since newborn mortality accounts for $57\%$ of total mortality in children under 5, reducing it has been challenging and sluggish in Bangladesh (\textcolor{blue}{Shams El Arifeen 2015}).
\begin{figure*}[ht]
\includegraphics[width=0.35\textwidth]{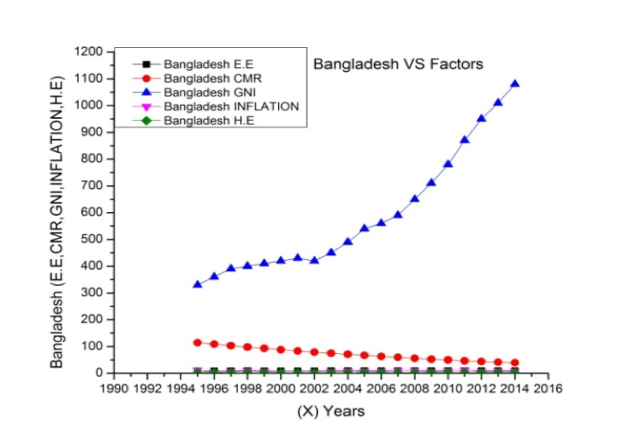}
\includegraphics[width=0.35\textwidth]{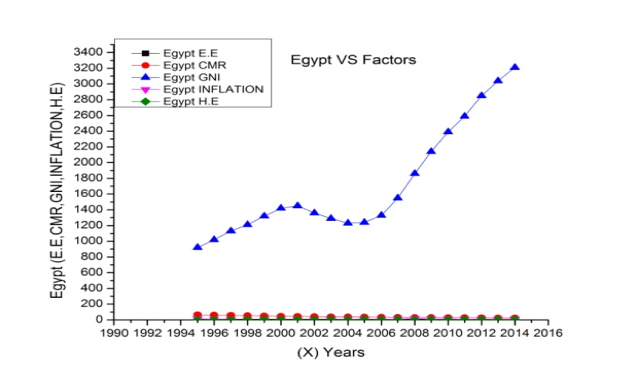}
\includegraphics[width=0.35\textwidth]{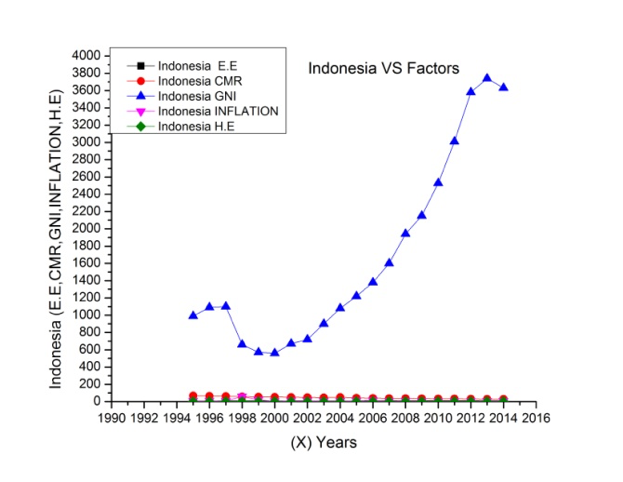}
\includegraphics[width=0.35\textwidth]{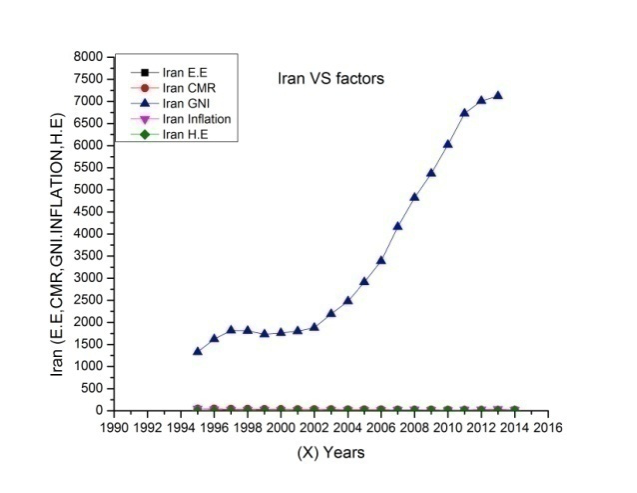}
\includegraphics[width=0.35\textwidth]{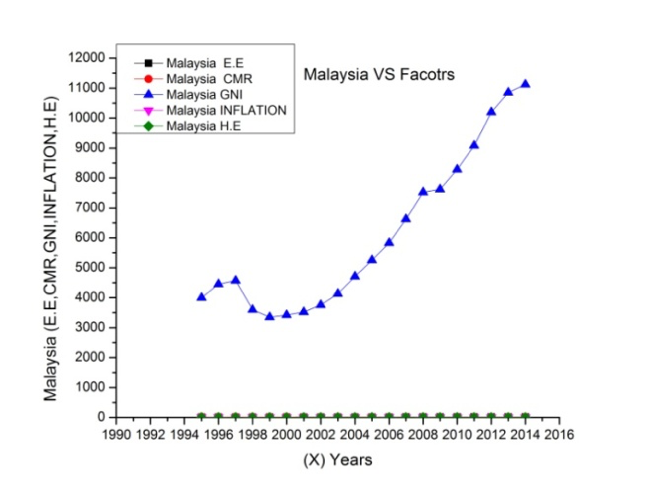}
\includegraphics[width=0.35\textwidth]{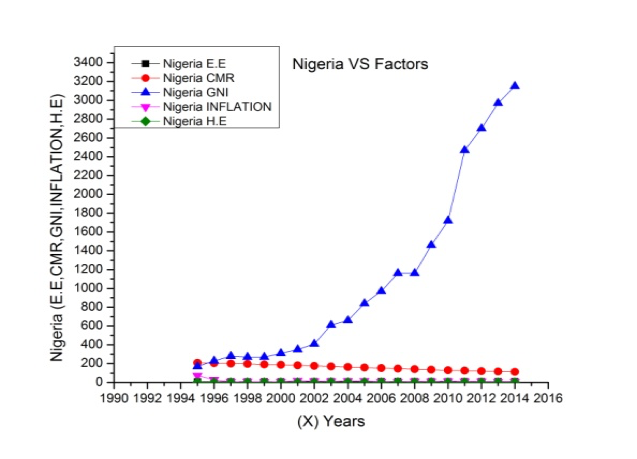}
\includegraphics[width=0.35\textwidth]{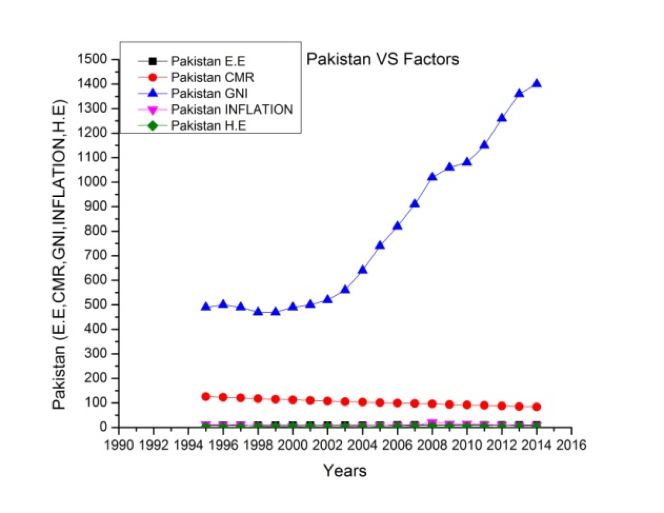}
\includegraphics[width=0.35\textwidth]{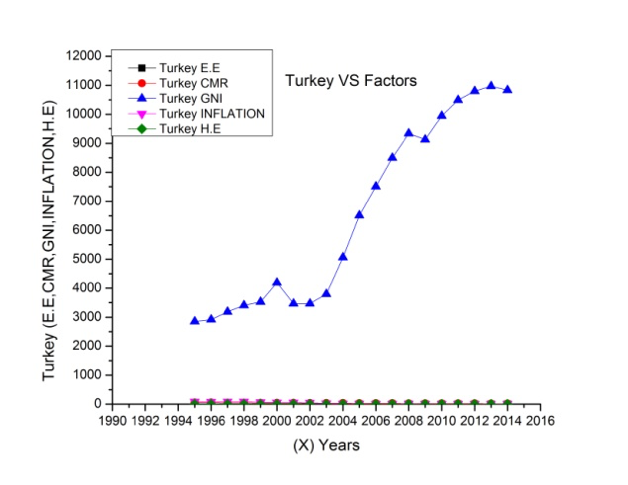}
\caption{Time-series trends of determinants (Health expenditures, education expenditures, GNI per capita, and inflation rate) influencing the CMU5 rate for each of the D-8 countries from 1995 to 2015.}\label{fig3}
\end{figure*}
\subsection*{(v) Inflation}
Inflation is an economic concept denoting a condition characterised by the general increase in prices of goods and services within a certain economy. As overall costs increase, customer purchasing power diminishes. The inflation rate denotes the measure of inflation over time. \textcolor{blue}{Figure 2D} illustrates the volatile inflation in D-8 nations, where abrupt price change policies have directly impacted infant mortality. To elucidate these variables and the D-8 nations, we generate additional graphs to analyse child mortality in each country as the dependent variable, with education spending, health expenditure, GNI per capita, and inflation serving as the independent factors. Time-series trends of determinants (Health expenditures, education expenditures, GNI per capita, and inflation rate) influencing the CMU5 rate for each of the D-8 countries from 1995 to 2015 are shown in \textcolor{blue}{Figure 3}.
\subsection*{(vi) D-8 countries vs Factors of Child Mortality}
Previously, we examined the component solely using D-8 nations data; nevertheless, we will now delineate the singular nations comprising all factors. This method involves the methodical identification of similar variables, along with their prevalence and rarity, individually. The x-axis denotes years, whereas the y-axis represents factors of D-8 as shown in \textcolor{blue}{Fig.3}. The aforementioned graphs demonstrate that GNI per capita is rising annually, whereas all other metrics remain static due to a deficiency of health expenditures in these D-8 nations.

All indicators, with the exception of GNI per capita, lack regulation by an authority of the D-8 nations. The comparison between Bangladesh and the factors \textcolor{blue}{Fig.3a} reveals that although GNI is rising at a steady rate, all other indicators, including GNI per capita, remain static. D-8 is functioning according to the established protocol. These aspects must improve respective sectors to more effectively address child mortality, hence promoting the development of Bangladesh. This protection is a crucial initiative to decrease child mortality. The graph for Egypt (\textcolor{blue}{Fig.3b}) demonstrates an increase in GNI, but all other indices are seeing a progressive decline. This form of reduction does not improve child death rates. Our finding indicates that Egypt similarly enhances its performance in these areas. Expenditure on education in D-8 nations is constant.
\subsection{Discussions }\label{sec3}
This study examined the determinants of child mortality in D-8 countries using panel data techniques. The panel unit root test results confirm that all variables are stationary at levels, ensuring the reliability of the estimated fixed- and random-effects models and ruling out spurious regression concerns.

The empirical findings consistently show that education expenditure plays a central role in reducing child mortality. Both fixed-effect and random-effect estimates reveal a large and statistically significant negative coefficient, indicating that increased public investment in education substantially improves child survival outcomes. This highlights the importance of education in enhancing health awareness, maternal knowledge, and access to preventive care.

GNI per capita also exhibits a significant negative relationship with child mortality, suggesting that higher income levels improve living standards, nutrition, and access to essential services. In contrast, inflation is found to increase child mortality, reflecting the adverse effects of rising prices on household purchasing power, food security, and healthcare affordability.

Interestingly, health expenditure does not show a statistically significant impact in either model. This may indicate inefficiencies in health spending allocation, unequal access to healthcare services, or governance and institutional constraints within D-8 countries. These findings suggest that merely increasing health budgets may not be sufficient without improvements in policy implementation and service delivery.

Overall, the results underscore that education and economic stability are more effective in reducing child mortality than health spending alone. Policymakers in D-8 countries should therefore prioritize education investment and macroeconomic stability alongside reforms aimed at improving the efficiency of health sector expenditures.

\section{Conclusion}\label{sec5}
This study analysed the effects of educational spending, health expenditure, GNI per capita, and inflation rate on under-five child mortality (U5MR) in D-8 nations from 1994 to 2014 by panel data analysis. Fixed- and random-effects models were analysed, and the Hausman test validated the superiority of the fixed-effects specification.

The findings demonstrate that educational investment is the primary factor influencing the reduction of child mortality, exhibiting a robust and consistent negative impact across all models. This underscores the essential function of education in enhancing child health outcomes via increased awareness, socioeconomic advancement, and access to resources. GNI per capita has a notable inverse correlation with child mortality, underscoring the significance of economic expansion in enhancing living standards and child survival rates.

Conversely, health expenditure does not have a statistically significant impact, indicating possible inefficiencies in the distribution of funds and the provision of services. Inflation is seen to elevate child mortality, indicating its detrimental effect on household purchasing power and access to vital products and healthcare services.

The data indicate that investment in education and wealth growth are more efficacious in decreasing child mortality than health expenditure alone in D-8 nations. Policy initiatives should thus prioritise the extension of education, the stabilisation of the economy, and the enhancement of efficiency in healthcare spending \textcolor{blue}{(Avelino et al. 2025)}.
\section*{Acknowledgments}
We extend our gratitude to M. Zubair (Concordia University, Canada), Ms. Noor-ul-ain (GSWCU Bahawalpur, Pakistan), Rao Ishtiaq Ahmed (Department of Economics, IUB, Bahawalpur, Pakistan), Ali Zaman (QAU, Islamabad), Ms. Abida (King's College London), and Ms. Fatima (London School of Economics) for their insightful discussions and valuable suggestions regarding our manuscript and data analysis. MWA expresses gratitude to the Director of Sports at The Islamia University Bahawalpur (IUB) for the undergraduate studentship and to the Chairman of the Department of Economics at IUB. MJA acknowledges the support provided by the Stocklin-Selmoni studentship at UCL. BS extends thanks to Tanveer Ahmed and Joana Loveday at Ravensbourne University for their insightful discussion and generous hospitality.
\appendix

\section*{References}

Adamou, H., Naba, G. and Koné, H., 2024. Socioeconomic inequalities in underweight children: a cross-sectional analysis of trends in West Africa over two decades. BMJ open, 14(2), p.e074522.
\\
\\
Adebayo, Samson B., and Ludwig Fahrmeir. "Analysing child mortality in Nigeria with geo-additive discrete‐time survival models." Statistics in Medicine 24.5 (2005): 709-728.
\\
\\
Ahmad, O.B., Lopez, A.D. and Inoue, M., 2000. The decline in child mortality: a reappraisal. Bulletin of the World Health Organization, 78, pp.1175-1191.
\\
\\
Asteriou, D. and Hall, S.G., 2021. Applied econometrics. Bloomsbury Publishing.
\\
\\
Athey, S. and Imbens, G.W., 2017. The state of applied econometrics: Causality and policy evaluation. Journal of Economic perspectives, 31(2), pp.3-32.
\\
\\
Avelino, I.C., Van-Dunem, J. and Varandas, L., 2025. Under-five mortality and social determinants in africa: a systematic review. European Journal of Pediatrics, 184(2), p.150.
\\
\\
Baltagi, B., 2008. Econometric analysis of panel data. John Wiley \& Sons.
\\
\\
Baltagi, B.H., 2008. Econometrics. Berlin, Heidelberg: Springer Berlin Heidelberg.
\\
\\
Balsalobre-Lorente, D., Nur, T. and Topaloglu, E.E., 2025. Examining the sustainable development process through the economic complexity, technology, urbanization, and renewable energy in D-8 countries. Quality \& Quantity, pp.1-41.
\\
\\
Baker, David P., et al. "The education effect on population health: a reassessment." Population and development review 37.2 (2011): 307-332.
\\
\\
Baker, D.P., Leon, J., Smith Greenaway, E.G., Collins, J. and Movit, M., 2011. The education effect on population health: a reassessment. Population and development review, 37(2), pp.307-332.
\\
\\
Behera, Manas Ranjan (2015). Relationship Between Maternal Education And Under-Five Mortality Rate In Low And Middle Income Countries-A Literature Review.  International Journal of Health Sciences and Research (IJHSR): 646-651.
\\
\\
Breitung, J., 2001. The local power of some unit root tests for panel data. In Nonstationary panels, panel cointegration, and dynamic panels (pp. 161-177). Emerald Group Publishing Limited.
\\
\\
Breitung, J. and Das, S., 2005. Panel unit root tests under cross‐sectional dependence. Statistica Neerlandica, 59(4), pp.414-433.
\\
\\
Black, Robert E., et al. "Global, regional, and national causes of child mortality in 2008: a systematic analysis." The lancet 375.9730 (2010): 1969-1987.
\\
\\
Bradshaw, J. ed., 2016. The Well-being of Children in the UK. Policy Press.
\\
\\
Baqui, Abdullah H., et al. "Effect of zinc supplementation started during diarrhoea on morbidity and mortality in Bangladeshi children: community randomised trial." Bmj 325.7372 (2002): 1059.
\\
\\
Basu, Alaka Malwade, and Rob Stephenson. "Low levels of maternal education and the proximate determinants of childhood mortality: a little learning is not a dangerous thing." Social science \& medicine 60.9 (2005): 2011-2023.
\\
\\
Blau, D. and van der Klaauw, W., 2007. The Impact of Social and Economic Policy on the Family Structure Experiences of Children in the United States. Department of Economics and Carolina Population Center, University of North Carolina, unpublished manuscript.
\\
\\
Bicego, George T., and J. Ties Boerma. "Maternal education and child survival: a comparative study of survey data from 17 countries." Social science \& medicine 36.9 (1993): 1207-1227.
\\
\\
Bokhari FAS, Gai Y, Gottret P. 2007. Government health expenditures and health outcomes. Health Econ, 16(3): 257–73.
\\
\\
Breitung, J., 2001. The local power of some unit root tests for panel data. In Nonstationary panels, panel cointegration, and dynamic panels (pp. 161-177). Emerald Group Publishing Limited.
\\
\\
Bryce, Jennifer, et al. "WHO estimates of the causes of death in children." The Lancet 365.9465 (2005): 1147-1152.
\\
\\
Cameron AC, Trivedi PK. 2005. Microeconometrics: methods and applications. Cambridge: Cambridge Press.
\\
\\
Campbell, J., Perron, P, 1991. Pitfalls and opportunities: what macroeconomists should know
about unit roots. In: Blanchard,O., Fischer,S. (Eds.), NBER Macroeconomics Annual. MIT Press, Cambridge, MA.
\\
\\
 Children: Reducing Mortality. World Health Organization (2016). Available online at: http://www.who.int/news-room/fact-sheets/detail/children-reducing-mortality (Accessed September 22, 2018).
 \\
 \\
  Childinfo.org: Statistics by Area – Child mortality – Under-five mortality. See $http://www.childinfo.org/mortality_underfive.php$ (accessed 7 March 2013)
\\ 
\\
Davies, A.; Lahiri, K. (1995). "A New Framework for Testing Rationality and Measuring Aggregate Shocks Using Panel Data". J. of Econometrics. 68 (1): 205–227.
\\
\\
Davidson, R., MacKinnon, J. G. 2004. Econometric Theory and Methods. New York: Oxford University Press. p. 613.
\\
\\
D’SOUZA, RENNIE M. "Role of health-seeking behavior in child mortality in the slums of Karachi, Pakistan." Journal of biosocial science 35.01 (2003): 131-144.
\\
\\
 El-Zanaty, Fatma, and Ann Adams Way. "Egypt Demographic and Health Survey 2000." (2006).
\\
\\
Ehret DY, Patterson JK, Bose CL. 2017.  Improving neonatal care: a global perspective. Clin Perinatol.
\\
\\
Frankenberg, Elizabeth. "The effects of access to health care on infant mortality in Indonesia." Health transition review (1995): 143-163.
\\
\\
Fayissa, B. and Traian, A., 2013. Estimation of a health production function: evidence from East-European countries. The American Economist, 58(2), pp.134-148.
\\
\\
Filmer D, Pritchett L. The impact of public spending on health: does money matter? Soc Sci Med. 1999;49(10):1309–23.
\\
\\
Fotio, H.K., Gouenet, R.M. and Ngo Tedga, P., 2024. Beyond the direct effect of economic growth on child mortality in Sub‐Saharan Africa: does environmental degradation matter?. Sustainable Development, 32(1), pp.588-607.
\\
\\
Goldman DP, Smith JP. Can patient self-management help explain the SES health gradient. Proc Natl Acad Sci. 2002;99:10929–10934. doi: 10.1073/pnas.162086599. 
\\
\\
 Hanmer, Lucia, Robert Lensink, and Howard White. "Infant and child mortality in developing countries: analysing the data for robust determinants." The Journal of Development Studies 40.1 (2003): 101-118.
\\
\\
Hatton, Timothy J. "Infant mortality and the health of survivors: Britain, 1910–50." The Economic history review 64.3 (2011): 951-972.
\\
\\
Hausman, J.A., 1978. Specification tests in econometrics. Econometrica: Journal of the econometric society, pp.1251-1271.
\\
\\
Hsiao, C.; Lahiri, K.; Lee, L.; et al., eds. (1999). Analysis of Panels and Limited Dependent Variable Models. Cambridge: Cambridge University Press.
\\
\\
Im, K.S., Pesaran, M.H. and Shin, Y., 2003. Testing for unit roots in heterogeneous panels. Journal of econometrics, 115(1), pp.53-74.
\\
\\
Iqbal, F. and Kiendrebeogo, Y., 2016. The determinants of child mortality reduction in the Middle East and North Africa. Middle East Development Journal, 8(2), pp.230-247.
\\
\\
Iqbal, F., Satti, M.I., Irshad, A. and Shah, M.A., 2023. Predictive analytics in smart healthcare for child mortality prediction using a machine learning approach. Open Life Sciences, 18(1), p.20220609.
\\
\\
Kim, K. and Moody, P.M., 1992. More resources better health? A cross-national perspective. Social science \& medicine, 34(8), pp.837-842.
\\
\\
Kravdal, Øystein. "Child mortality in India: the community-level effect of education." Population studies 58.2 (2004): 177-192.
\\
\\
Karunarathne, M., Buddhika, P., Priyamantha, A., Mayogya, P., Jayathilaka, R. and Dayapathirana, N., 2025. Restoring life expectancy in low-income countries: the combined impact of COVID-19, health expenditure, GDP, and child mortality. BMC Public Health, 25(1), p.894.
\\
\\
Khalid, A.M., Ferguson, R.J. and Asadullah, M.N. eds., 2023. Economic integration among D-8 Muslim countries: Prospects and challenges.
\\
\\
Latif, A.A., 2025. D-8 member states: Economic profiles, sectors and opportunities. Synergies, Cooperation and Trajectaries, p.12.
\\
\\
 Liang L, Kotadia N, English LL, Kissoon N, Kabakyenga J, Ansermino JM, Lavoie PM, Wiens MO. Predictors of Mortality in Neonates and Infants Hospitalized With Sepsis or Serious Infections in Developing Countries: A Systematic Review. Frontiers in pediatrics. 2018;6:277.
 \\
 \\
 Liu L, Johnson HL, Cousens S, et al. Global, regional, and national causes of child mortality: an updated systematic analysis for 2010 with time trends since 2000. Lancet 2010; 379: 2151–61. [PubMed]
\\
\\
Liu, Li, et al. "Global, regional, and national causes of child mortality in 2000–13, with projections to inform post-2015 priorities: an updated systematic analysis." The Lancet 385.9966 (2015): 430-440.
\\
\\

Lo Bue, M.C., 2024. Drivers of changes in child nutritional conditions: A panel data‐based study on Indonesian households, 1997–2014. Review of Development Economics, 28(2), pp.741-776.
\\
\\
Maddala, G. S. (2001). Introduction to Econometrics (Third ed.). New York: Wiley. 
\\
\\
Mathews, T. J., and Marian F. MacDorman. "Infant mortality statistics from the 2003 period linked birth/infant death data set." National vital statistics reports 54.16 (2006): 1-29.
\\
\\
Martin, Linda G., et al. "Co-variates of child mortality in the Philippines, Indonesia, and Pakistan: an analysis based on hazard models." Population Studies 37.3 (1983): 417-432.
\\
\\
Maruthappu, M., Ng, K.Y.B., Williams, C., Atun, R. and Zeltner, T., 2015. Government health care spending and child mortality. Pediatrics, pp.peds-2014.
\\
\\
Mejia, S.A., 2024. Globalization, foreign direct investment, and child mortality: A cross-national analysis of less-developed countries, 1990–2019. International Journal of Comparative Sociology, 65(3), pp.378-406.
\\
\\
Miller, Jane E., et al. "Birth spacing and child mortality in Bangladesh and the Philippines." Demography 29.2 (1992): 305-318.
\\
\\
Mirowsky, J., 2017. Education, social status, and health. Routledge.
\\
\\
Muhamad, I.A.K., Rifdah, B.N., Hidhayad, A.P. and Kusdiwanggo, S., 2024, April. Grasping the essence of the Millennium Development Goals: A literature review. In IOP Conference Series: Earth and Environmental Science (Vol. 1324, No. 1, p. 012059). IOP Publishing.
\\
\\
Nourin, S., Nasim, I., Raza ur Rehman, H.M., Montero, E.C., Garat de Marin, M.S., Abdel Samee, N. and Ashraf, I., 2025. Exploring the nexus: Hausman test application in tourism, globalization, and environmental sustainability-evidence of top 10 visited countries. Humanities and Social Sciences Communications, 12(1), pp.1-12.
\\
\\
Novignon, J., Olakojo, S.A. and Nonvignon, J., 2012. The effects of public and private health care expenditure on health status in sub-Saharan Africa: new evidence from panel data analysis. Health Economics Review, 2(1), p.22.
\\
\\
Novignon J, Nonvignon J, Mussa R, Chiwaula L. Health and vulnerability to poverty in Ghana: evidence from the Ghana Living Standards Survey Round 5. Health Economics Review. 2012.
\\
\\
O'Hare, B., Makuta, I., Chiwaula, L. and Bar-Zeev, N., 2013. Income and child mortality in developing countries: a systematic review and meta-analysis. Journal of the Royal Society of Medicine, 106(10), pp.408-414.
\\
\\
O'Hare, B. and Makuta, I., 2015. An analysis of the potential for achieving the fourth millennium development goal in SSA with domestic resources. Globalization and health, 11(1), p.8.
\\
\\
Pamuk, Elsie R., Regina Fuchs, and Wolfgang Lutz. "Comparing relative effects of education and economic resources on infant mortality in developing countries." Population and Development Review 37.4 (2011): 637-664.
\\
\\
Pandey, Manoj K. "Maternal health and child mortality in rural India." (2009).
\\
\\
Pearson, E.S., 1931. The test of significance for the correlation coefficient. Journal of the American Statistical Association, 26(174), pp.128-134.
\\
\\
Pelletier, David L., et al. "The effects of malnutrition on child mortality in developing countries." Bulletin of the World Health Organization 73.4 (1995): 443.
\\
\\
Peterson, Andrew. "Introduction." Compassion and Education. Palgrave Macmillan UK, 2017. 1-12.
\\
\\
Phillips, P.C.B., 1988. Testing for a Unit Root in Time Series Regression. Biometrika.
\\
\\
Preston S. The changing relation between mortality and level of economic development. Population Studies 1975; 29: 231–48 , $http://www.jstor.org/discover/10.2307/2173509?uid=2\&uid=4\&sid=21102308701567$ [PubMed]
\\
\\
Pritchett L, Summers L. 1994. “Wealthier is Healthier”. J Hum Resour, 31:4.
\\
\\
P. Musgrove. “Public and Private Roles in Health,” 1996. [Online]. Available:
http://siteresources.worldbank.org/
HEALTHNUTRITIONANDPOPULATION/Resources/281627-
1095698140167/Musgrove-PublicPrivate-whole.pdf.
19th September 2015.
\\
\\
Rahman, M.M. and Khanam, R., 2017. Health care expenditure and health outcome nexus: new evidence from SAARC-ASEAN region.
\\
\\
Rigby, Michael J., et al. "Child health indicators for Europe: a priority for a caring society." The European Journal of Public Health 13.suppl 3 (2003): 38-46.
\\
\\
 Rajaratnam JK, Marcus JR, Flaxman AD, Wang H, Levin-Rector A, Dwyer L, et al. Neonatal, postneonatal, childhood, and under-5 mortality for 187 countries, 1970–2010: a systematic analysis of progress towards millennium development goal 4. Lancet (2010) 375:1988–2008. doi: 10.1016/S0140-6736(10)60703-9
 \\
 \\
Sarkar, Dipanwita, and Jayanta Sarkar. "Persistence of income inequality: does child mortality matter?." The Journal of Developing Areas 46.2 (2012): 105-123.
\\
\\
Selcuk, M. and Asutay, M., 2025. Assessing financial convergence in developing countries: The case of D-8 countries. Borsa Istanbul Review.
\\
\\
Shah, S.Z., Chughtai, S. and Simonetti, B., 2020. Renewable energy, institutional stability, environment and economic growth nexus of D-8 countries. Energy Strategy Reviews, 29, p.100484.
\\
\\
Shandra, C. L., Shandra, J. M., \& London, B. (2011). World Bank structural adjustment, water, and sanitation: A cross-national analysis of child mortality in sub-Saharan Africa. Organization \& Environment, 24(2), 107.
\\
\\
Shetty, Anil, and Shraddha Shetty. "The Impact of Female Literacy on Infant Mortality Rate in Indian States." Current Pediatrics 18.1 (2014).
\\
\\
Smith-Greenaway, Emily. "Maternal reading skills and child mortality in Nigeria: a reassessment of why education matters." Demography 50.5 (2013): 1551-1561.
\\
\\
Song, Shige, and Sarah A. Burgard. "Dynamics of inequality: mother’s education and infant mortality in China, 1970-2001." Journal of Health and Social Behavior 52.3 (2011): 349-364.
\\
\\
Summers L, Pritchett L. Wealthier is healthier. World Bank Policy Research Working Paper No. 1150. Washington, DC: World Bank; 1993.
\\
\\
United Nations. The Millennium Development Goals Report 2014. 2014.
\\
\\
United Nations. The millenium development goals report. New York: UN; 2010.
\\
\\
Unicef. Committing to child survival: a promise renewed. eSocialSciences, 2015.
\\
\\
Van der Klaauw, Bas, and Limin Wang. "Child mortality in rural India." Journal of Population Economics 24.2 (2011): 601-628.
Gage, Timothy B., et al. "Maternal education, birth weight, and infant mortality in the United States." Demography 50.2 (2013): 615-635.
\\
\\
Verbeek, M., 2008. A guide to modern econometrics. John Wiley \& Sons.
\\
\\
Victora CG, Barros AJ, Axelson H, et al. How changes in coverage affect equity in maternal and child health interventions in 35 Countdown to 2015 countries: an analysis of national surveys. Lancet. 2012;380(9848):1149–1156pmid:22999433
\\
\\
Wagstaff, Adam. "Poverty and health sector inequalities." Bulletin of the world health organization 80.2 (2002): 97-105.
\\
\\
Wang L. Health outcomes in low-income countries and policy implications: empirical findings from demographic and health surveys. World Bank Policy Research Working Paper No. 2831. Washington, DC: World Bank; 2002;1–35.
\\
\\
Wang, H., Liddell, C.A., Coates, M.M., Mooney, M.D., Levitz, C.E., Schumacher, A.E., Apfel, H., Iannarone, M., Phillips, B., Lofgren, K.T. and Sandar, L., 2014. Global, regional, and national levels of neonatal, infant, and under-5 mortality during 1990–2013: a systematic analysis for the Global Burden of Disease Study 2013. The Lancet, 384(9947), pp.957-979.
\\
\\
Wang, H., Bhutta, Z.A., Coates, M.M., Coggeshall, M., Dandona, L., Diallo, K., Franca, E.B., Fraser, M., Fullman, N., Gething, P.W. and Hay, S.I., 2016. Global, regional, national, and selected subnational levels of stillbirths, neonatal, infant, and under-5 mortality, 1980–2015: a systematic analysis for the Global Burden of Disease Study 2015. The Lancet, 388(10053), pp.1725-1774.
\\
\\
Wang, H., Abajobir, A.A., Abate, K.H., Abbafati, C., Abbas, K.M., Abd-Allah, F., Abera, S.F., Abraha, H.N., Abu-Raddad, L.J., Abu-Rmeileh, N.M. and Adedeji, I.A., 2017. Global, regional, and national under-5 mortality, adult mortality, age-specific mortality, and life expectancy, 1970–2016: a systematic analysis for the Global Burden of Disease Study 2016. The Lancet, 390(10100), pp.1084-1150.
\\
\\
Wang, F., Gillani, S., Balsalobre‐Lorente, D., Shafiq, M.N. and Khan, K.D., 2025. Environmental degradation in South Asia: Implications for child health and the role of institutional quality and globalization. Sustainable Development, 33(1), pp.399-415.
\\
\\
Ward, I.L., Barrett, S.L., Razieh, C., Standeven, C., Zylbersztejn, A., John, E., Zaccardi, F., Modi, N., Khunti, K., Ayoubkhani, D. and Nafilyan, V., 2025. Maternal ethnic group, socioeconomic status, and neonatal and child mortality: a nationwide cohort study in England and Wales. The Lancet Public Health, 10(9), pp.e774-e783.
\\
\\
Ware Jr, John E. "SF-36 health survey update." Spine 25.24 (2000): 3130-3139.
\\
\\
Ware, Helen. "Effects of maternal education, women's roles, and child care on child mortality." Population and Development Review 10 (1984): 191-214.
\\
\\
WB. Book The World development indicators. The World Bank; 2012. The World development indicators.
\\
\\
WHO. World health statistics. Geneva: World Health Organization; 2010.
\\
\\
World Health Organization. Global tuberculosis report 2013. World Health Organization, 2013.
\\
\\
World Health Organization. MDG 4: reduce child mortality. Geneva, Switzerland: WHO; 2013. 
\\ Available at: http://www.who.int/topics/millennium \\ \_development\_goals/child\_mortality/en/. Accessed May 5, 2014
\\
\\
You D, Hug L, Ejdemyr S. Global regional, and national levels and trends in under-5 mortality between 1990 and 2015, with scenario-based projections to 2030: a systematic analysis by the UN Inter-agency Group for child mortality estimation. Lancet (2015) 386:2275–86. doi: 10.1016/S0140-6736(15)00120-8
\\
\\
You, Danzhen, New Rou Jin, and Tessa Wardlaw. "Levels \& trends in child mortality." (2012).
\\
\\
Zaman, A., 2025. Negating Time and History: A Metaphysical Defense of Ahistorical Liberalism. Available at SSRN 5992794.
\\
\\
Zupan J. Perinatal mortality in developing countries. N Engl J Med. (2005) 352:2047–8. doi: 10.1056/NEJMp058032

\bibliography{waseem-arxiv}
\end{document}